\documentstyle[psfig]{mn}

\def\porb{P_{\rm orb}}
\def\pspin{P_{\rm spin}}
\def\pbeat{P_{\rm beat}}
\def\rco{R_{\rm co}}
\def\rcirc{R_{\rm circ}}
\def\be{\begin{equation}}
\def\ee{\end{equation}} 

\begin{document}

\title{The Spin Period of EX Hydrae}
\author[A.R. King \& G.A. Wynn]{A.R. King \& G.A. Wynn,\\ Astronomy Group, 
University of Leicester, Leicester, LE1~7RH}

\newcommand{\lta}{{\small\raisebox{-0.6ex}{$\,\stackrel
{\raisebox{-.2ex}{$\textstyle <$}}{\sim}\,$}}}
\newcommand{\gta}{{\small\raisebox{-0.6ex}{$\,\stackrel
{\raisebox{-.2ex}{$\textstyle >$}}{\sim}\,$}}}   

\maketitle
\begin{abstract}
We show that the spin period of the white dwarf in the magnetic CV 
EX~Hydrae represents an equilibrium
state in which the corotation radius is comparable with the distance
from the white dwarf to the inner Lagrange point. We also show that 
a continuum of spin equilibria exists at which $\pspin$ is
significantly longer than $\sim 0.1 \porb$. Most systems occupying these 
equilibrium states should have orbital periods below the CV period
gap, as observed. 
\end{abstract}

\section{Introduction}
\label{sec:intro}
Cataclysmic variables (CVs) are semi--detached binaries in which a
low--mass, main--sequence secondary star transfers mass to a white
dwarf. In a significant number of systems the white dwarf has a
magnetic moment strong enough to channel the accretion flow on to
restricted regions of its surface; such systems are known as magnetic
CVs. Observationally these systems are characterized by the presence of
two coherent periods $\porb, \pspin$, corresponding to the orbital
rotation of the binary and the white dwarf spin respectively. Sometimes
the beat period 
\begin{equation}
\pbeat = (\pspin^{-1} - \porb^{-1})^{-1}
\label{1}
\end{equation}
is seen in addition to or even in place of $\pspin$. Magnetic CVs are
interesting because the white dwarf spin rates give insight into the
angular momentum flows within the binary. These flows are more complex than
in the analogous systems (pulsing X--ray binaries) in which the
accretor is a magnetic neutron star,
simply because the white dwarf magnetic moments $\mu_1$  ($ \gta
10^{33}$~G cm$^3$) are much larger than those of the neutron stars
($\mu_1 \lta 10^{30}$~G cm$^3$). 
Accordingly the magnetospheric radius in a magnetic CV is a
significant fraction of the binary separation $a$, whereas it is far
smaller in the neutron--star case. As an illustration of this, the
white dwarf spin can be locked to the binary rotation ($\pspin =
\porb$) in some magnetic CVs (called AM Her stars), while $\pspin$ is
always considerably smaller than $\porb$ in neutron--star
systems. 

Historically the neutron--star systems were discovered first,
and simple theories of the angular momentum flows were developed for
them (e.g. Ghosh \& Lamb, 1979). There has been a natural tendency to
transfer these theories bodily to the white dwarf case, even though the
larger magnetic moments make this very questionable. In particular,
neutron--star magnetospheres are so small that Roche lobe overflow must
always lead to the formation of a Keplerian accretion disc in the standard
way (e.g. Frank et al., 1992). By contrast, a white dwarf magnetic
field can have a strong influence on the accretion flow at
all radii, making it non--Keplerian even though it may surround the
white dwarf. This in turn tends to lead to an equilibrium spin rate
determined approximately by the condition (King, 1993; Wynn \& King, 1995)
\begin{equation}
\rco \sim \rcirc.
\label{2}
\end{equation}
Here
\be
\rco = \biggl({GM_1\pspin^2\over 4\pi^2}\biggr)^{1/3}
\label{3}
\ee
is the corotation radius, at which matter in 
local Kepler rotation about a white dwarf of mass $M_1$ corotates
with the magnetic field lines, and $\rcirc$ is the circularization
radius, i.e. the radius at which the Kepler specific angular momentum
equals that of matter accreting through the inner Lagrange point
$L_1$. If the latter is a distance $b$ from the white dwarf we can
find an approximation for $\rcirc$ by equating the specific angular
momentum $\sim b^2(2\pi/\porb)$ at $L_1$ to the specific Kepler
angular momentum at $\rcirc$ (this neglects the effects of the companion) 
\be 
\rcirc \simeq {M\over M_1}
\biggl({b\over a}\biggr)^4\biggl({GM\porb^2\over 4\pi^2}\biggr)^{1/3},
\label{4}
\ee
where $M=M_1+M_2$ is the total binary mass. Numerical fits to Roche
geometry give
\be
{b\over a} = 0.500 - 0.227 \log {M_2\over M_1}
\label{5}
\ee
(Warner 1976). Hence combining (\ref{2} - \ref{5}) gives a spin
equilibrium 
\be
{\pspin\over \porb} \sim 
\biggl({b\over a}\biggr)^6\biggl({M\over M_1}\biggr)^2\sim 0.07
\label{6}
\ee
for typical mass ratios (cf King, 1993, Wynn \& King, 1995). The numerical
simulations performed by Wynn \& King (1995) used the magnetic drag 
prescription introduced by King (1993). Here the flow is
assumed to be in the form
of diamagnetic blobs, the local magnetic field $B =
\mu_1/r^3$ acting on them through a surface drag. This produces an acceleration
\begin{equation}
{\bf f}_{\rm mag} = -k [{\bf v} - {\bf v}_{\rm f}]_\bot, 
\label{9}
\end{equation}
where ${\bf v}$ and 
${\bf v}_{\rm f}$ are the blob and field velocities (i.e. $v_{\rm f} =
2\pi r/\pspin$), the suffix $\bot$ 
denotes the velocity components perpendicular to the field lines, and
$k = t_{\rm mag}^{-1}$, with $t_{\rm mag}$ the drag time scale given by
\begin{equation}
t_{\rm mag} \simeq \frac{c_{A}\rho_{\rm b}l_{\rm b}}{B^{2}}  
\label{10}
\end{equation}
(Drell, Foley and Ruderman, 1965). Here $\rho_{\rm b}, l_{\rm b}$
are the blob density and length scale, and
$c_{\rm A}$ is the Alfv\'en speed in the inter-blob plasma.
Thus as the blobs move
through the magnetosphere they exchange orbital energy
and angular momentum with the white dwarf via the drag term. The blob
parameters $\rho_{\rm b}, l_{\rm b}$ are to some extent arbitrary, but
limited for example by physical conditions near $L_1$. Numerical
simulation of the flow including the full Roche potential rather than
the simple approximation (\ref{4}) examined the regime in which 
\be 
t_{\rm mag}(\rcirc) \sim t_{\rm dyn}(\rcirc),
\label{tm1}
\ee
or alternatively
\be 
t_{\rm mag}(L_1) \gg t_{\rm dyn}(L_1),
\label{tm2}
\ee
where $t_{\rm dyn} \sim (r^3/GM_1)^{1/2}$ is the dynamical time scale.
These simulations give similar but slightly larger values of 
$\pspin / \porb$ to the approximation (\ref{6}).

Equation (\ref{6}) reasonably describes the relation between
$\pspin$ and $\porb$ in most non--synchronous magnetic CVs
(`intermediate polars', or IPs). The corresponding accretion flow
 is sometimes referred to as
`discless'. This is a rather misleading name, suggesting for
example that the flow cannot surround the white
dwarf, although simulations show that this is possible (Fig. 1). 
Observations suggesting for example that matter is fed
to the white dwarf at all spin phases are sometimes taken as evidence
against this type of accretion. However the fundamental property of this flow
is not the absence of matter surrounding the white dwarf, but simply
the non--Keplerian nature of the velocity field. This in turn reflects
the pervasive influence of the magnetic field on the accretion
dynamics. The divergence from Keplerian flow is a manifestation of the complex
angular momentum flow within the system,
where the angular momenta of the blobs may be transferred to the primary 
star via the field or accretion, recaptured by the secondary star
or lost to the binary entirely. This contrasts with the case of a
Keplerian accretion disc where the angular momentum transferred
through $L_1$ is passed back to the binary orbit via tidal forces
at the outer edge of the disc.

Since the IPs cover a large range of magnetic field and orbital period
there are several exceptions to the equilibrium represented by
(\ref{6}). One would expect
systems with Keplerian discs to have $\rco < \rcirc$ and thus a
smaller ratio $\pspin/\porb$ than given by (\ref{6}). This appears to
be the case in at least two systems. In GK~Per
($\pspin = 381$ s, $\porb = 48$ hr) the orbital period is so long that
the magnetospheric radius is small compared with $\rcirc$, so a
Keplerian disc will form in the standard way. In DQ~Her ($\pspin =
71$~s, $\porb = 4.65$~hr) the very short spin period presumably
indicates a rather low magnetic field, leading to the same
conclusion. However, the still shorter spin period of AE~Aqr ($\pspin
= 33$~s, $\porb = 9.88$~hr) does {\it not} imply a Keplerian disc
(Wynn, King \& Horne, 1997, and references therein); here the matter
transferred from the secondary star is apparently being centrifugally
ejected, showing that disc formation depends on $\pspin$ as well as
the magnetic moment $\mu_1$ (cf. equation 7). 

In addition to the systems with $\pspin/\porb$ smaller than implied by
(\ref{6}) there are two 
IPs with a much larger ratio: EX~Hya has $\pspin =
67$~min, $\porb = 98$~min, and RX1238-38 has $\pspin = 36$~min, 
$\porb = 90$~min. This clearly does not fit the `IP'
spin equilibrium (\ref{6}), still less the equilibrium expected
for accretion from a Keplerian disc. A spin period far from
equilibrium would imply that we are seeing these systems in a very
short--lived phase: in the case of EX~Hya the spin period derivative
is $\dot{P}_{\rm spin} = - 3.8\times 10^{-11}$ s s$^{-1}$ (eg. Warner, 1995),
with a spindown timescale of $\sim 10^6$ y, much shorter than the
orbital evolution timescale ($\gta 10^8$ y).
It is far more likely that the spin is currently close
to equilibrium, but that this equilibrium differs from that
represented by (\ref{6}) or the Keplerian case. In this paper we shall
show that EX~Hya represents a new type of spin equilibrium for IPs
with short orbital periods.

\section{Spin Equilibrium in EX Hydrae}
\label{sec:spin}
Most IPs so far discovered have orbital periods on the long side of
the CV period gap 2~hr $\lta \porb \lta \; 3$~hr. EX~Hya's short orbital
period is therefore very suggestive: if its magnetic field is similar
to systems conforming to (\ref{6}) we would expect the smaller orbital
separation and lower mass transfer rate
to make the effective magnetospheric radius comparable with
the distance to the secondary star. Here magnetic forces will dominate
the motion of the gas blobs near $L_1$ and 
\be 
t_{\rm mag}(L_1) \lta t_{\rm dyn}(L_1).
\label{11}
\ee
Setting $\rho_b$ and $l_b$ to be equal to the expected stream density
($\sim 10^{-9}$ g cm$^{-3}$) and width ($\sim 10^9$ cm) at $L_1$
respectively,
we find $t_{\rm mag}(L_1) \sim t_{\rm dyn}(L_1)$ for 
$\mu_1(\rm crit) \sim 3 \times 10^{33}$ G cm$^3$ (cf fig. 6). 
For magnetic moments larger
than this we might then expect to see the
equilibrium condition (\ref{2}) (which leads to eq. \ref{6}) replaced
by
\be
\rco \sim b,
\label{7}
\ee
leading (from \ref{3}, \ref{5}) to 
\be
{\pspin\over \porb} \sim \biggl({b\over a}\biggr)^{3/2} =
\biggr(0.500-0.227\log {M_2\over M_1}\biggl)^{3/2}.
\label{8}
\ee
For the mass ratio $M_2/M_1 = 0.19$  (Hellier,
1996) appropriate to EX Hya we find $\pspin/\porb \sim 0.54$, similar to
the measured $\pspin/\porb = 0.68$. However this simple treatment
completely neglects the full Roche potential. Since the angular momentum
exchange determining $\pspin/\porb$ in this type of flow occurs near
$L_1$ we expect a significant modification of this result because of
the presence of the secondary star (this effect was less serious for
the IP equilibrium (\ref{6}) as the angular momentum exchange there occurs
much closer to the white dwarf). We therefore need numerical
simulations using the full Roche potential. 

Clearly the condition (\ref{11}) favours larger
values of the magnetic moment $\mu_1$. However this must not be so
large that the system synchronizes and becomes an AM Her system. This
condition can be expressed as (e.g. King, Frank \& Whitehurst, 1991)
\be
{\mu_1\mu_2\over a^3} < \dot {2\pi M b^2\over \porb},
\label{12}
\ee
where $\mu_2$ is the secondary dipole and $\dot M$ is the mass
transfer rate. In practice this condition requires 
\be
\mu_1 \lta 10^{34} \; {\rm G \: cm}^3.
\label{13}
\ee
Spin evolution calculations similar to those of Wynn and King (1995) 
show that the spin period of the white dwarf in EX~Hya reaches
an equilibrium which is in excellent agreement with the 
observed $\pspin/\porb = 0.68$ ratio, for $t_{\rm mag}(L_1)
/ t_{\rm dyn}(L_1) \lta 1/3$ (see section 3 and figure 5). 
Spin equilibrium was attained on a time-scale $t_{\rm spin} \sim 
10^7$ yrs, much shorter than the orbital evolution time-scale. 
This numerical result contrasts with the simple estimate above
which predicted a ratio of $\pspin/\porb = 0.54$ from the white 
dwarf potential alone. We conclude that (\ref{7}) indeed specifies 
a spin equilibrium state, and that EX Hya is close to this state.

Figure 2 shows a sequence of views of the accretion flow in the
equilibrium spin state of our model system. This suggests that
accretion of mass and angular momentum by the
white dwarf occurs only for a restricted range of spin phases,
essentially when the dipole axis points towards $L_1$. At other phases
the condition (\ref{7}) shows that matter trying to accrete along
field lines will be centrifugally ejected. Much of this ejecta will be
re--accreted by the secondary, transferring angular
momentum from the white dwarf back to the binary orbit. In equilibrium
these angular momentum flows balance over each orbit. We note here
that recent spectroscopic observations (Wynn, Wheatley and Maxted,
1999) show strong evidence for just this sort of flow pattern in 
EX~Hya.
Figure 3 shows the
fractions of mass
transferred instantaneously through $L_1$ which are accreted, returned
to the companion star, or ejected entirely from the system for the 
simulation depicted in figure 2. It can clearly be seen that most of
the transferred mass is accreted by the white dwarf, and this
accretion occurs in two 'bursts' each beat cycle. It should be noted
however, that the simulation takes no account of the modulation of the
accretion flux by the white dwarf spin which may well dominate the 
observed light curve.

The spin equilibrium described here is quite robust and, in
particular, may be achieved with a large spread in blob parameters:
only $\sim 25\%$ of the accreting mass need be magnetically dominated
(i.e. condition \ref{11} holds) to achieve the EX~Hya equilibrium.
Figure 4 shows the results of a simulation with a greater spread in
blob parameters. In this figure it is clear that the accretion flow
consists of 2 components, one of which is similar to that described
above and the other is a azimuthally symmetric distribution of
particles surrounding the white dwarf. This complex system could mimic
many of the observational properties expected of accretion via a disc
and via a stream, whilst still attaining an equilibrium $\pspin/\porb$
ratio very close to that observed for EX~Hya. 

\section{A Continuum of Spin Equilibria}

The equilibrium conditions (\ref{2}) and (\ref{7}) suggest that there
exists a continuum of equilibrium states in which
\be
\rcirc \lta \rco \lta b.
\label{con3}
\ee
Figure 5 shows this set of equilibrium states for various values of 
$t_{\rm mag}(L_1)/t_{\rm dyn}(L_1)$, in the
case of a binary with system parameters identical to those of EX~Hya.
Using (\ref{10}) to provide an estimate of $\mu_1$ from $t_{\rm
mag}(L_1)$ it can be seen that
the equilibrium states between the usual IP equilibrium 
\be
\pspin / \porb \sim 0.1, \;\;\;
5 \times 10^{30} \; {\rm G \: cm^3} \lta \mu_1 \lta 10^{31} \; {\rm G \: cm^3} 
\ee
and the EX~Hya equilibrium
\be
\pspin / \porb \sim 0.68, \;\;\;
10^{33} \; {\rm G \: cm^3} \lta \mu_1 \lta 10^{34} \; {\rm G \: cm^3} 
\ee
(with the exact value of each depending upon the mass ratio of the
system) are defined by 
\be
0.1 \lta \pspin / \porb \lta 0.68, \;\;\;
10^{31} \; {\rm G \: cm^3} \lta \mu_1 \lta 10^{33} \; {\rm G \: cm^3}. 
\ee
For $\mu_1 \lta 5 \times 10^{30}$ G cm$^3$ a standard Keplerian
accretion disc would be expected to form, and for 
$\mu_1 \gta 10^{34}$ G cm$^3$ the system would synchronize and be 
observable as an AM~Her star. Primary magnetic moments $\lta 10^{31}$
G cm$^3$ are much lower than those normally inferred for IPs above the
period gap ($\gta 10^{32}$ G cm$^3$). Therefore, we would expect IPs
above the period gap to evolve into EX~Hya-like systems or AM Hers.
It should be noted that each system would have a unique curve similar
to that presented in figure 5, which would be a function of $\porb$,
$M_2/M_1$ and $\dot{M}$.

The spin period of the white dwarf in the 
short period magnetic CV RX1238-38 ($\pspin / \porb \simeq 0.4$)
may represent one of these continuum equilibrium states. 
Confirmation of this is dependent on the (as yet undetermined) 
mass ratio of the system. However, the
$\pspin / \porb$ ratio is clearly too high to represent the usual IP 
equilibrium \ref{6}. If $\pspin$ was determined by $\rco \sim b$ 
then (\ref{8}) gives a rough estimate of the mass ratio of the system 
as $M_2 / M_1 \sim 0.65$. This mass ratio is rather high for a system 
below the period gap and the most likely possibility is that RX1238-38
represents one of the possible spin equilibria defined by (\ref{con3}).

\section{Discussion}
\label{sec:dis}

\noindent
The reasoning above shows that EX Hya represents a new
equilibrium spin state for magnetic CVs characterized by the
approximate equality (\ref{7}). We can now show that the necessary
conditions (\ref{11}, \ref{13}) generally confine this equilibrium
state to short orbital
periods, in precise agreement with observation. In Figure 6 we plot 
$t_{\rm mag}(L_1)/t_{\rm dyn}(L_1)$ versus $\porb$ for three values
of $\mu_1$. (We have assumed secular mean values for $M_2, \dot M$, cf
e.g. King, 1988.) Evidently the required conditions for the EX Hya
equilibrium hold only for a narrow range 
\be
2\times 10^{33} \; {\rm \: G cm}^3
\lta \mu_1 \lta 10^{34} \; {\rm \: G cm}^3,
\label{14}
\ee
(in agreement with Figure 6), and then only for 
\be
\porb < 2\ {\rm hr},
\label{15}
\ee
i.e. periods below the gap. For (\ref{11}) to hold at periods
above the gap would require values of $\mu_1$ probably violating (\ref{13}),
i.e. such systems are generally synchronous (AM Her systems). Thus we expect
most systems in this equilibrium state to be below the period gap.

There is a continuum of equilibrium spin states which connect the usual
IP equilibrium and that represented by EX~Hya. The spin equilibrium
for a given $\porb$, $M_2/M_1$ and $\dot{M}$ is determined 
by the strength of the primary's magnetic moment. The other short period
magnetic CV RX1238-38 most likely occupies one of these states. 
The relatively low values of $\mu_1$ ($\gta 10^{31}$ g cm$^3$)
required to attain such equilibria at short orbital periods ($\lta 2$ hr)
indicates that magnetic CVs above the orbital period gap will evolve
toward long spin periods (either $\pspin > 0.1 \porb$, or $\pspin =
\porb$) below the gap. It is
possible that some of these longer spin equilibria may be attainable
above the period gap. 

We may now crudely classify magnetic CVs according to $\porb$ and
$\mu_1$ (Fig. 9). We see that EX Hya--type systems have
field strengths similar to typical IPs above the
period gap and comparable with the weakest--field AM Her systems below
it. Systems with larger $\mu_1$ are evidently AM Her systems
below the gap even if not above it. The Figure prompts the obvious
question as to why there are so few asynchronous magnetic CVs below 
the period gap. There are a number of possibilities:
\begin{enumerate}
\item the field distribution has a lower bound, and such systems are
rare. DQ Her would be an example above the period gap.
\item Such systems do exist, but are difficult to detect, particularly below 
the period gap.
\item Orbital angular momentum losses associated with the ejected
material drive the orbital evolution of short period systems
significantly more quickly than gravitational radiation alone. 
\end{enumerate}
In the second case, we should remember that IPs are usually found
through their medium--energy ($\gta 2$ keV) emission. The lower mass
transfer rates below the period gap may simply make this too weak for
detection. 

We have shown that the spin period of EX Hya 
corresponds to a new equilibrium state in which the corotation radius
is comparable to the distance to the inner Lagrange point. Moreover
this equilibrium is only likely below the period gap, because the
magnetic fields required above the gap would make the systems
synchronous. The accretion flow (Fig. 1) characterizing this state
has distinctive features which should allow observational
tests. We note that Wynn, Wheatley and Maxted (1999) present data 
which provide strong evidence for exactly this form of accretion/ejection 
flow in EX~Hya.  

\section{Acknowledgments}
Research in theoretical astrophysics at the University of Leicester is
supported by a PPARC rolling grant. ARK gratefully acknowledges a
PPARC Senior Fellowship.

\bigskip
\noindent
{\bf Figure Captions}\\
\newline
\noindent
{\bf Figure 1} {Numerical simulations of the accretion flow in EX~Hya, adopting the
magnetic drag prescription of King (1993), and a dipolar magnetic field
geometry. The upper panels depict the
system as viewed from above the orbital (x-y) plane, where the motion of the
secondary is in a anti-clockwise sense and the z-axis is orientated out
of the plane of the paper. The lower panels depict the system as
viewed from the x-z plane. In each case a half-Gaussian distribution
for $k$ was used: to reflect a stream of blobs with a dense
core and tenuous wings (higher $k$ values). The distribution mean 
and standard deviation were $t{\rm mag}(L_1)/t_{\rm dyn}(L_1) = 0.25$ and 
$0.025$ (tailing to lower values) respectively. The left-hand panels
show the results of a simulation which also includes an equal 
number of particles with very low $k$ values,
in order to demonstrate the robust nature of the spin equilibrium 
(see section 2).}
\newline
\noindent
{\bf Figure 2} {A sequence of views of the accretion flow in the
equilibrium state for a simulation with magnetic parameters equal to
those of the right-hand panels of figure 1. The left hand panels in
the figure show the x-y projection of the system, while the right-hand
panels show the corresponding x-z projection. Each set of panels 
are separated in time by 0.2 $\porb$.}\\
\newline
\noindent
{\bf Figure 3} {Fractions of mass transferred through $L_1$ which are
accreted by the white dwarf (solid curve), recaptured by the secondary
star (dot-dash curve), and escaping from the binary (dashed curve),
for the simulation shown in figure 2.}\\
\newline
\noindent
{\bf Figure 4} {As figure 2, but for a simulation with magnetic
parameters equal to those of the simulation shown in the left-hand
panels of figure 1.}\\
\newline
\noindent
{\bf Figure 5} {Equilibrium $\pspin/\porb$ values as a function of
$t_{\rm mag}(L_1)/t_{\rm dyn}(L_1)$, and estimated $\mu_1$ for
EX~Hya.}\\
\newline
\noindent
{\bf Figure 6} {$t_{\rm mag}(L_1)/t_{\rm dyn}(L_1)$
verses $\porb$. From top to bottom the panels plot
the time-scales for $\mu_1 = 10^{33}$, $2 \times 10^{33}$, 
$4 \times 10^{33}$ G cm$^3$.}\\
\newline
\noindent
{\bf Figure 7} {Schematic classification of magnetic CVs according to 
$\mu_1$ and $\porb$. Here DQ~Her systems are taken to represent systems
which accrete via a Keplerian disc and have $\pspin/\porb \ll 0.07$.}\\

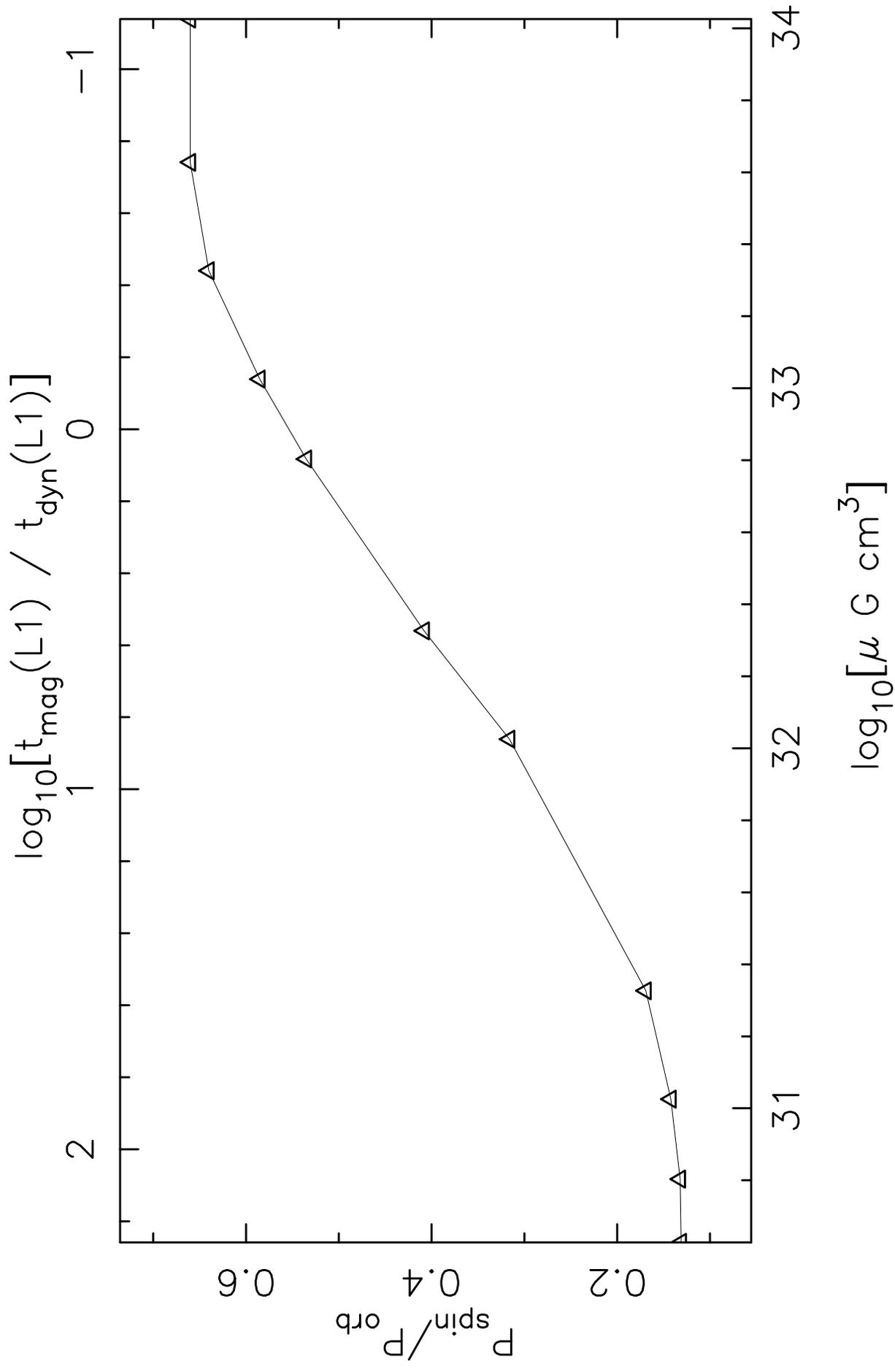
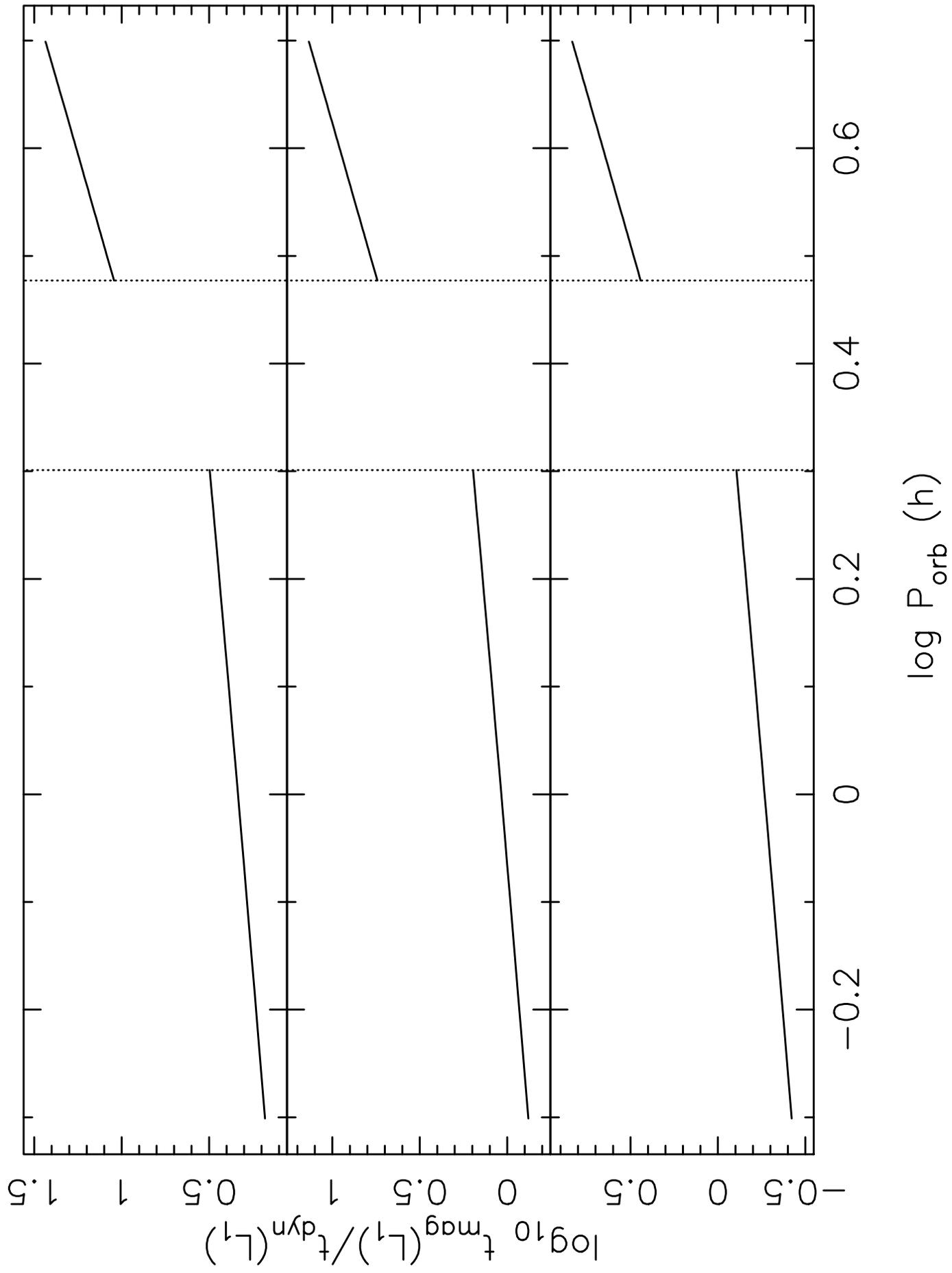
\end{document}